\documentclass[amsmath,amssymb,aps,prl,twocolumn,superscriptaddress]{revtex4-2}

\usepackage{latexsym}
\usepackage{amssymb}
\usepackage{mathrsfs}
\usepackage{amsmath}
\usepackage{soul}
\usepackage{bm}
\usepackage{verbatim}
\usepackage{epsfig}
\usepackage{multirow}
\usepackage{xspace}
\usepackage{ltablex}
\usepackage{ulem}
\usepackage[english]{babel}
\usepackage[utf8x]{inputenc}
\usepackage[T1]{fontenc}
\usepackage[usenames,dvipsnames]{xcolor}

\usepackage{color}
\definecolor{darkblue}{rgb}{0,0,0.6}
\definecolor{darkred}{rgb}{0.6,0,0}

\usepackage{hyperref}
\hypersetup{
	bookmarksopen=true,
	pdftitle="",
	pdfauthor="",
	pdfsubject="",
	pdfstartview={FitH},
	pdfmenubar=true,
	pdfhighlight=/O,
	colorlinks=true,
	urlcolor=darkblue,
	citecolor=darkblue,
	linkcolor=MidnightBlue,
}

\usepackage{cleveref}

\begin{document}
	
%%\title{Estimate the Density of Two-level systems in glasses}
\title{Toward understanding the depletion of two-level systems in ultrastable glasses}
\author{Wencheng~Ji}

\affiliation{%
	Institute of Physics,
	\'Ecole Polytechnique F\'ed\'erale de Lausanne (EPFL),
	CH-1015 Lausanne,
	Switzerland}

\affiliation{Department of Physics of Complex Systems, Weizmann Institute of Science, Rehovot 76100, Israel}
\affiliation{School of Engineering and Applied Sciences, Harvard University, Cambridge, Massachusetts 02138, USA}

\begin{abstract}
 The density of Two-level systems (TLS)  controls the low-temperature thermal properties in glasses and has been found to be almost depleted in ultrastable glasses.   While  this depletion of TLS is thought to have a close relationship with the dramatic decrease of quasi-localized modes (QLMs), it has yet to be clearly formalized.  
 In this work, we argue, based on the \textit{soft-potential} model, that TLS correspond to QLMs with typical frequency  $\omega_0$.
 The density $n_0$ of TLS is proportional to both the density of QLMs  $D_L(\omega_0)$, and the fraction of symmetric double-wells $f(\omega_0)$ at $\omega_0$, i.e., $n_0 \propto D_L(\omega_0)f(\omega_0)$.
 We numerically estimate $\omega_0$ and $n_0$ in computer glasses at different levels of stabilities, and find that $\omega_0$ is about $5\%$ to $10\%$ of the Debye frequency.  
 $n_0$ in ultrastable glasses is over $1000$ times smaller than that in poorly prepared glasses,  with both $D_L(\omega_0)$ and $f(\omega_0)$ decreasing significantly. Remarkably,  the order of magnitude of estimations for $n_0$ agrees with that found in experiments in amorphous silicon. Our study paves the way to understanding  the depletion of TLS through the rarefaction of QLMs. 
 
\end{abstract}
\maketitle
\renewcommand{\floatpagefraction}{.5}
\paragraph{Introduction:}
In glasses, an atom or a group of atoms can tunnel between two nearby energy wells, which corresponds to the switch between two energy states called two-level systems (TLS)  \cite{Phillips87}.
 Although the entities of TLS are still not clear, the  phenomenological \textit{tunneling} TLS model proposed by Anderson et al.  \cite{Anderson72}  and Phillips \cite{Phillips72} successfully explains the  abnormal thermal properties in glasses around $1$ Kelvin ($K$), whose specific heat is approximately  linear in temperature $T$ and thermal conductivity  almost proportional to $T^2$  \cite{Zeller71}.  In many previous studies, the density of TLS  $n_0$ has been found to be nearly constant \cite{Phillips81,Freeman86,Berret88,Pohl02}.  However, thanks to the state-of-art vapor deposition method by which ultrastable glasses can be prepared in experiments, the specific heat  around $1\,K$  in such glasses is found to be proportional to $T^3$ \cite{Perez14, Liu14}. This finding implies that $n_0$ decreases significantly  in ultrastable glasses \cite{Ramos20}, which is also supported by the latest numerical simulations where  $n_0$ decreases by a factor of $100$ \cite{Khomenko20}.  However, $n_0$ in these simulations is still two orders of magnitude greater than the experimental results.  From a theoretical perspective, the role of interactions between TLS, which influences $n_0$, is still widely debated \cite{Yu88, Parshin07, Lubchenko07, Faoro15}. On the practical aspect, it is essential to reduce $n_0$  because TLS cause decoherence in superconducting qubits which are a promising candidate for the construction of quantum computers \cite{Martinis05, Arute19,Muller19}.

$n_0$ is related to the concentration of double-well potentials (DWPs), and the latter is thought to be linked to the density of local vibrational modes, namely, quasi-localized modes (QLMs) since DWPs have large spatial overlaps with some fraction of QLMs \cite{ Parshin07, Wencheng19,Khomenko21}.  
QLMs (normal modes of the Hessian) are present at low-frequency vibrational spectrum \cite{Schober93} and can be easily obtained nowadays in computer glasses. 
In most glasses, the density of QLMs has the form $D_L(\omega)=A_4\omega^4$ for small frequency $\omega$ \cite{Baity15,Lerner16,Mizuno17,Lerner18,Shimada18,Wang19,Rainone20,Lerner21}. The prefactor $A_4$ is also found to decrease by several hundreds of times in ultrastable glasses compared to poorly prepared glasses \cite{Rainone20}. However, the origin of QLMs, specifically the dramatic decrease of $A_4$ is debated  \cite{Wencheng20,Rainone21}. Moreover, the quantitative relationship between $A_4$ and $n_0$ is unclear.

\begin{figure}[hbt!]
	\centering
	\includegraphics[width=\linewidth]{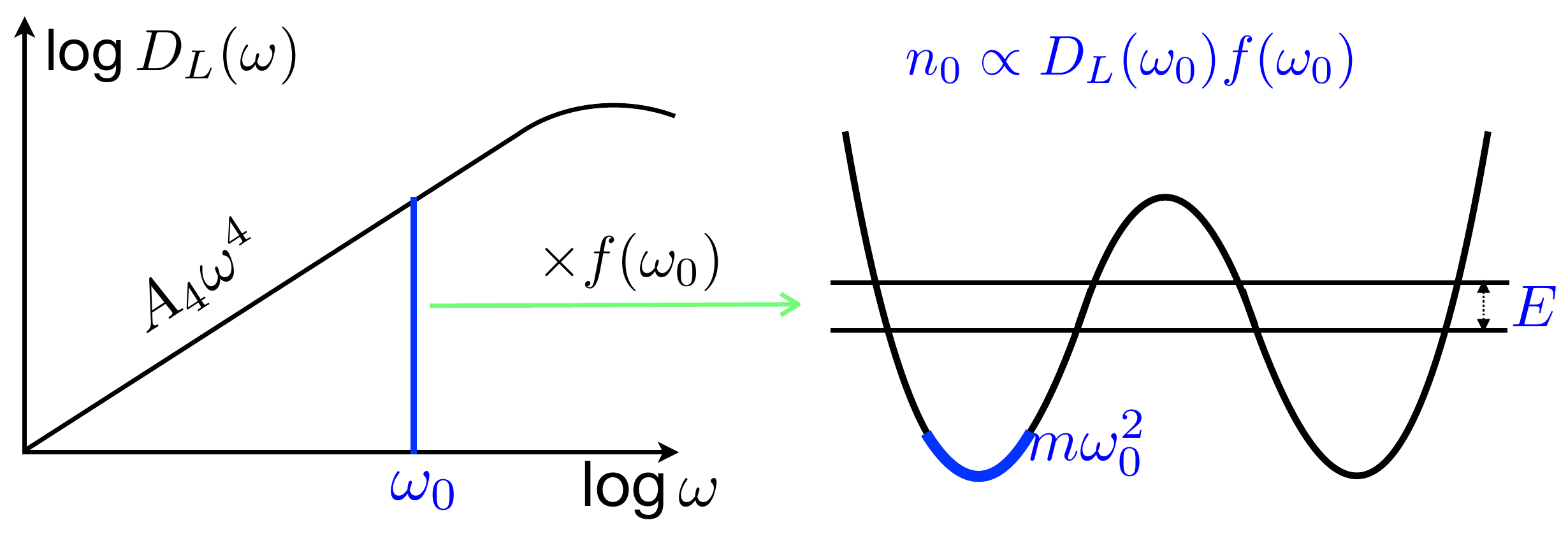}
	\caption{Sketch of the relationship between the density of TLS $n_0$ and QLMs $D_L(\omega)$.  $D_L(\omega)=A_4\omega^4$ for small $\omega$ (left); symmetric  DPWs along the reaction coordinate  with the energy splitting $E\sim k_BK$,  which corresponds to a near-constant curvature $m\omega_0^2$ at minima for a given system preparation (right), where $m$ is the mass of the particle and $k_B$ is the Boltzmann constant.  Since the reaction coordinate that is curvilinear is along the direction of QLMs around the minimum \cite{Khomenko21}, TLS correspond to the QLMs with frequency $\omega_0$.
	Their density $n_0 \propto D_L(\omega_0) f(\omega_0)$ where  $f(\omega_0)$ reflects the fraction of symmetric DWPs. 
}
	\label{fig:sketch}
\end{figure}

  In this paper, we argue, based on the \textit{soft-potential} model, that  TLS correspond to the QLMs with typical frequency $\omega_0$. Their density $n_0$  is proportional to  both $D_L(\omega_0)$ and the fraction of symmetric  DWPs $f(\omega_0)$ at $\omega_0$, as shown in Fig.~\ref{fig:sketch} with a sketch.  We numerically estimate $\omega_0$ and $n_0$ in computer glasses at different levels of stabilities.  We find that $\omega_0$ is about $5\%$ to $10\%$  of the Debye frequency, and  $n_0$ in ultrastable glasses is more than $1000$ times smaller than poorly prepared glasses, with both $D_L(\omega_0)$ and $f(\omega_0)$ playing a significant role. Remarkably, the order of magnitude of estimations for $n_0$ agrees with that found in experiments in amorphous silicon.

\paragraph {Model:}

We follow the spirit of the \textit{soft-potential} model \cite{Ilyin87,Galperin89,Parshin94} by which the local potential of double wells along the reaction coordinate $s$ can be written approximately to the fourth-order of the Taylor expansion around a minimum. This local potential reads
\begin{align}
U(s) & =\frac{1}{2}m\omega^{2}s^{2}+\frac{\kappa}{3!}s^{3}+\frac{1}{4!}\chi s^{4},
\label{U}
\end{align}
where $m$ is the mass of a particle with equal mass.  We assume that the curvilinear reaction coordinate \footnote{The reaction coordinate that connects two well is the path of minimum energy.} is along the direction of QLMs around the minimum, which is supported by Ref.~\cite{Khomenko21}. $\omega$ thus corresponds to an eigenfrequency of the QLMs.  Physically, $\chi>0$ which ensures that the potential has a lower energy limit.  Due to the numerical observation that $\chi$ is narrowly distributed \cite{Lerner16}, $\chi$ is set to be a constant for a given system preparation and becomes larger in more stable glasses \cite{Wencheng21}. $\kappa$ is set to be positive  without loss of generality \footnote{ If $\kappa<0$, flip the sign of it by using $-s$ without changing $\omega^2$ and $\chi$.}. The joint density distribution $P(\omega,\kappa)$ is given by 
\begin{align}
P(\omega,\kappa)=\frac{1}{3N}\sum_{i}\sum_{j}\text{\ensuremath{\delta(\omega-\omega_{i})}\ensuremath{\delta(\kappa-\kappa_{j})}}.
\end{align}
where $N$ is the number of particles in a three-dimensional glass, $\omega_i$ and $\kappa_j$ are the discrete values of $\omega$ and $\kappa$.

In the quantum tunneling regime, the DWPs are of near equal depth which indicates
$\kappa\approx\kappa_{c}\equiv\omega \sqrt{3m\chi}$. According to the \textit{tunneling} TLS model \cite{Phillips72,Anderson72}, the energy splitting $E  =(\left(\delta\varepsilon\right)^{2}+ \Delta_0^{2})^{1/2}$ where $\delta\varepsilon$
is energy asymmetry and $\Delta_0$ is the  tunneling contribution  derived from the WKB approximation. $\Delta_0$ reads
\begin{equation}
	\Delta_{0} \approx W\text{\ensuremath{\exp}\ensuremath{\left[-\intop_{s_{{\rm 1}}}^{s_{2}}\frac{\sqrt{2mU(s)}}{\hbar}\,ds\right]}}=W\text{\ensuremath{\exp}\ensuremath{\left[-\left(\frac{\omega}{\bar{\omega}}\right)^{3}\right]}}
	\label{Delta}
\end{equation}
where $s_{1,2}=-2\kappa_c/\chi,\,0$, are two
minima of the symmetric double well. $\bar{\omega}=\left(\hbar\chi/\left(2m^{2}\right)\right)^{1/3}$. $W$ is a characteristic energy
and we adopt $W=\hbar\left(\hbar\chi/\left(96m^{2}\right)\right)^{1/3}$
same with the choice in  Ref.~\cite{ Parshin93}.  Note that $\Delta_{0}$ is insensitive to the change of $W$ compared to $\omega$.

Now we can obtain the joint density distribution $P(E,\Delta_0)$ in the vicinity of symmetric wells from $P(\omega,\kappa)$ where $\kappa=\kappa_c+\delta\kappa$ and $\delta\kappa \ll \kappa_c$.
\begin{align}
& P\left(E,\Delta_0\right)=P(\omega,\kappa)\left\Vert \frac{d\omega d\kappa}{dEd\Delta_0}\right\Vert \approx P_0\frac{E}{\Delta_0\sqrt{E^{2}-\Delta_0^{2}}}
\label{P_ED}
\end{align}
where
\begin{align}
P_0=\frac{P(\omega,\kappa_{c})\kappa_{c}}{18\hbar}\left(\frac{\bar{\omega}}{\omega}\right)^{6}.
\end{align}
In the above calculation (see detail in Support Information (SI)), we considered the leading order of $\kappa$ around $\kappa_{c}$, that is $P(\omega,\kappa)\approx P(\omega,\kappa_c)$ where $P(\omega,\kappa_c)$ indicates the density of modes with symmetric wells. 

$\Delta_0$ has a lower bound $\Delta_{\min} \approx 10^{-13} k_BK$. It corresponds to the experimental time scale $\tau \sim 100 \rm{s}$ \footnote{$\Delta_{\rm min}$ is estimated by $\Delta_{\rm min}=\hbar\pi/\tau$}. At large $\Delta_0$ (small $\omega$), $P_0$ would be proportional to $\left[\ln(W/\Delta_0)\right]^{-2/3}$  since $P(\omega, \kappa_c)$ is argued to be proportional to $\omega^3$ \cite{Parshin93,Wencheng19}  \footnote{ At very tiny $\Delta_0$, $P(\omega, \kappa_c)$ might be near-constant or  might (dramatically) decrease by increasing $\omega$.}. Since $P_0$ decreases with decreasing $\Delta_0$, it weakens the contributions of $P\left(E,\Delta_0\right)$ at tiny $\Delta_0$ compared to regarding $P_0$ as a constant. The lower bound $\Delta_{\min}$  would play a less significant role than the upper bound $E$. For example, from a simple numerical calculation, $\int_{\Delta_{\min}}^{10\Delta_{\min}}P(E,\Delta_{0})d\Delta_{0}$
is about $6$ times smaller than $\int_{0.1E}^{E}P(E,\Delta_{0})d\Delta_{0}$
that is about $24\%$ of the total integral value, where we use $W=10E$. The former integral could be further smaller if $P(\omega, \kappa_c)$ is saturated at tiny $\Delta_0$ (large $\omega$).   From Eq.~(\ref{Delta}),  $\omega$ is narrowly distributed \footnote{When $W=10E$,  $\omega$ is $2.4$ times larger at $\Delta_{\min}$than that at $\Delta_0=E$}. Hence,  taking all the above factors into account, we think that TLS correspond to a typical frequency for a given system preparation:
\begin{align}
\omega_0=\left[\ln\left(\frac{W}{E}\right)\right]^{1/3}\bar{\omega},
\label{w0}
\end{align}
which is obtained by setting to $\Delta_0=E$ in Eq.~(\ref{Delta}). From the above equation, $\omega_0$ is insensitive to the  magnitude of $E$ where $E \sim k_BT$ at low temperature \cite{Phillips87}.  The existing typical $\omega_0$ is verified in the recent numerical work \cite{Khomenko21}. 

To integrate $\Delta_0$ out in Eq.~(\ref{P_ED}) to get $P(E)$, we set to $\Delta_0 = E$ in $P_0$ since it varies slowly compared to $1/(\Delta_0\sqrt{E^{2}-\Delta_0^{2}})$ and will not change the order of magnitude of the integral \footnote{If we adopt $P_0\propto \left[\ln(W/\Delta_0)\right]^{-2/3}$, then the integral is approximately $3$ times smaller when $W=10E$.}.  Also note that the conventional definition of the density of TLS $n_0(E)\equiv3NP(E)/V$  is the number of TLS per volume per energy \cite{Phillips81}. At last, $n_0(E)$ reads
\begin{align}
n_0(E)\approx \frac{P(\omega_0,\kappa_0)\kappa_0}{6\hbar a^3} \ln^{-2}\left(\frac{W}{E}\right)\ln\left(\frac{2E}{\Delta_{\textrm{min}}}\right),
\label{PE}
\end{align}
where $\kappa_0=\omega_0\sqrt{3m\chi}$ and $a\equiv (V/N)^{1/3}$ is the inter-particle distance. Considering that $P(\omega_0,\kappa_0)\kappa_0\propto \left[\ln(W/E)\right]^{4/3}$ also depends on $E$, $n_0(E)$ varies slowly with varying $E$. Hence, from this place, we set to $E=1k_BK$ \footnote{If $E=10k_BK$, $P(E)$ only increases by around $70\%$ compared to $E=k_BK$ where we use $W=100\,k_BK$. } to estimate the density of TLS $n_0(E=k_BK)$ ($n_0$ for short):
 \begin{align}
	n_{0}= C_0 P(\omega_0,\kappa_0)
	%n_{0}(E=k_BK) \equiv \frac{3NP(E=k_BK)}{V} \approx C_0 P(\omega_0,\kappa_0)
	\kappa_0,
	\label{n0}
\end{align}
where $C_{0}\approx\frac{31}{6\hbar a^{3}}\ln^{-2}\left(W/ k_BK\right)$. The value of $\Delta_{\min}$ is assigned.  For a given material, $C_0$ is insensitive to the system preparation since $a$ and the logarithmic term slightly change.  Therefore the change of the magnitude of $n_0$ is determined by $P(\omega_0,\kappa_0)$.  Once we know the value of $P(\omega_0,\kappa_0)$, we can get $n_0$. However, to get the reaction coordinates, also called minimum energy paths (MEPs), is computationally very expensive \cite{Khomenko21}. We estimate $P(\omega_0,\kappa_0)\kappa_0$ in another simple way.  

Qualitatively, $P(\omega_0,\kappa_0)\kappa_{0}$ is directly related to the density of QLMs  $D_L(\omega_0)$.  But only the fraction of QLMs $f(\omega_0)$ correspond to DWPs with symmetric wells since QLMs also correspond to single wells and for DWPs just a fraction of them are symmetric DWPs.  To build up the relation, we rewrite
\begin{align}
	\ensuremath{P(\omega_{0},\kappa_{0})\kappa_{0}}=D_{L}(\omega_{0})f(\omega_{0}),
	\label{pwk}
\end{align}
where 
\begin{align}
	f(\omega_{0})& \equiv\frac{P\left(\omega_0,\kappa_{0}\right)\kappa_{0}}{\int_{0}^{\infty}P\left(\omega_0,\kappa\right)d\kappa}.
	\label{f_w}
\end{align}
Note that $D_{L}(\omega)\equiv\int_0^\infty P(\omega,\kappa)d\kappa =\frac{1}{3 N} \sum_{i} \delta\left(\omega-\omega_{i}\right)$, which includes both single wells and double wells \footnote{Since single wells do not correspond to reaction coordinates, this part of integral can be calculated  along directions of QLMs in Eq.(\ref{f_w}). It will not change the value of the integral. Of course, if we know that which QLMs correspond to DWPs, we know minimum energy paths and directly use Eq.~(\ref{n0}).}.  We will see later that $D_{L}(\omega_0)$ can be calculated exactly and dimensionless $f(\omega_0)$ can be estimated by a way. 

We have no restriction on the form of $D_L(\omega_0)$ with respect to $\omega_0$ up to now. In a regular glass, $D_L(\omega_0)=A_4\omega_0^4$ still holds at $\omega_0$ (see the discussion in next section). Therefore, 
  \begin{align}
 n_0 \propto A_4 
 \end{align}
 It explains why $n_0$ has a dramatic decrease with decreasing parent temperatures \cite{Khomenko20} since $A_4$ dramatically decreases with it as well below some typical temperature \cite{Wang19, Rainone20}.   If there are no double wells (only single wells),  regardless of the magnitude of $D_L(\omega_0)$,  $n_0=0$ because of $f(\omega_{0})=0$.  Hence, both $D_L(\omega_0)$ and $f(\omega_{0})$ influence the magnitude of $n_0$ \footnote{In \cite{Pricaccia21}, the authors think that the density of QLMs for the part of DPWs is proportional to $\omega^3$ which is incompatible with our consideration if $D_L(\omega)\sim \omega^4$ for small $\omega$, which is argued in our previous work \cite{Wencheng19}.}.   
\\

%If $D_L(\omega)$ is gapped in some glasses\cite{Kapteijns19,Wencheng19b} and the gap magnitude $\omega_c$ is larger than $\omega_0$,  then $n_0$ would decrease significantly.
% \footnote{In \cite{Wencheng19b}, we have a typical frequency $\omega_c^*$ below which TLS exist. It is a necessary condition. In our approach here, $\omega_0<\omega_c^*$}.

\paragraph{Numerical estimation:}

We estimate the values of $n_0$ in three-dimensional zero-temperature  computer glasses that are quenched instantaneously from equilibrated configurations at four parent temperatures $T_p=0.8,\,0.55,\,0.35,\,0.3$.  The glass transition temperature $T_g$ is about $0.5$. See the model detail in \cite{Lerner19}.  Here, ten thousand equilibrated configurations at each $T_p$ ($T_p=0.55,\,0.35,\,0.3$) at $N=2000$ prepared by the SWAP Monte Carlo method \cite{Ninarello17}  are taken from \cite{Rainone20}.  Equilibrated configurations with $1000$ realizations at $T_p=0.8$ at $N=8000$ are prepared by the normal molecular dynamics method directly. 

 The typical frequency $\omega_0$ depends on  $W$ and $\bar{\omega}$ both of which are as a function of $\chi$.  We let $\chi=c_{1}m\omega_{D}^{2}/a^{2}$  where $\omega_D$ is the Debye frequency. Dimensionless $c_1$ is estimated numerically.  The typical frequency $\omega_0$ obtained from  Eq.~(\ref{w0}) can be expressed as  
\begin{align}
\frac{\omega_0}{\omega_D}=\left[\frac{c_{1}\hbar}{2m\omega_{D}a^{2}}\ln\left(\frac{\hbar\omega_{D}}{k_{B}K}\left(\frac{c_{1}\hbar}{96m\omega_{D}a^{2}}\right)^\frac{1}{3} \right)\right]^\frac{1}{3}.
\label{w0_estimate}
\end{align}
$\omega_0$ is approximately proportional to $c_1^{1/3}$.  We know that $m\omega_{D}^{2}a^{2}$ is an order of the banding energy $10\,\text{eV}$ and $\hbar\omega_{D}\sim50 \,\text{meV}$ because the Debye temperature is usually several hundred Kelvin.  If $(c_1/2)^{1/3}$ is of order $1$,  as a rough estimation, $\omega_0\sim 0.1\omega_D$. 

To do the careful calculations of $\omega_0$, $D_L(\omega_0)$ and $f(\omega_0)$, we explicitly adopt the parameters of the amorphous silicon where $m=4.8\times10^{-26}\,\text{kg}$ and $a=2.9\times10^{-10}\,\text{m}$ \cite{Queen13}. $\omega_D$ at different $T_p$ is slightly different because the elastic moduli increase with decreasing $T_p$. We set to $\omega_D=530k_{B}K/\hbar$ \cite{Mertig84} at $T_p=0.55$ since it is close to the glass transition temperature. See SI how we estimate $\omega_D(T_p)$ at the other three $T_p$.   Specifically, $c_1\approx 0.5,\,0.8,\,2.1,\,2.5$ from high to low $T_p$, which is taken from the median of $\chi_1/(m\omega_{D}^{2}/a^{2})$ (see the definition of $\chi_1$ later).  As a result, $\omega_0/\omega_D \approx 0.06,\,0.07,\,0.94,\,0.10$ (listed in TABLE~\ref{tabn0}.) is consistent with our rough estimations above.  $\omega_0/\bar{\omega}$  are  about $1.24,\,1.27,\,1.33,\,1.34$, which is less insensitive to the system preparation as we expect from Eq.~(\ref{w0}). 

In such computer glasses, the density of QLMs has a pseudogap at low frequency $D_{L}(\omega) =A_{4} \omega^{4}$. 
The upper bound of this scaling regime is about $20\%$ to $30\%$ of $\omega_D$ \cite{Rainone20}.  So the quartic spectrum of QLMs at $\omega_0$ holds although they would be hybridized with the plane waves. We extract $A_4$ from $D_L(\omega)$ (see SI) for small $\omega$ and then get $D_L(\omega_0)=A_4\omega_0^4$.

\begin{figure}[hbt!]
	\centering
	\includegraphics[width=\linewidth]{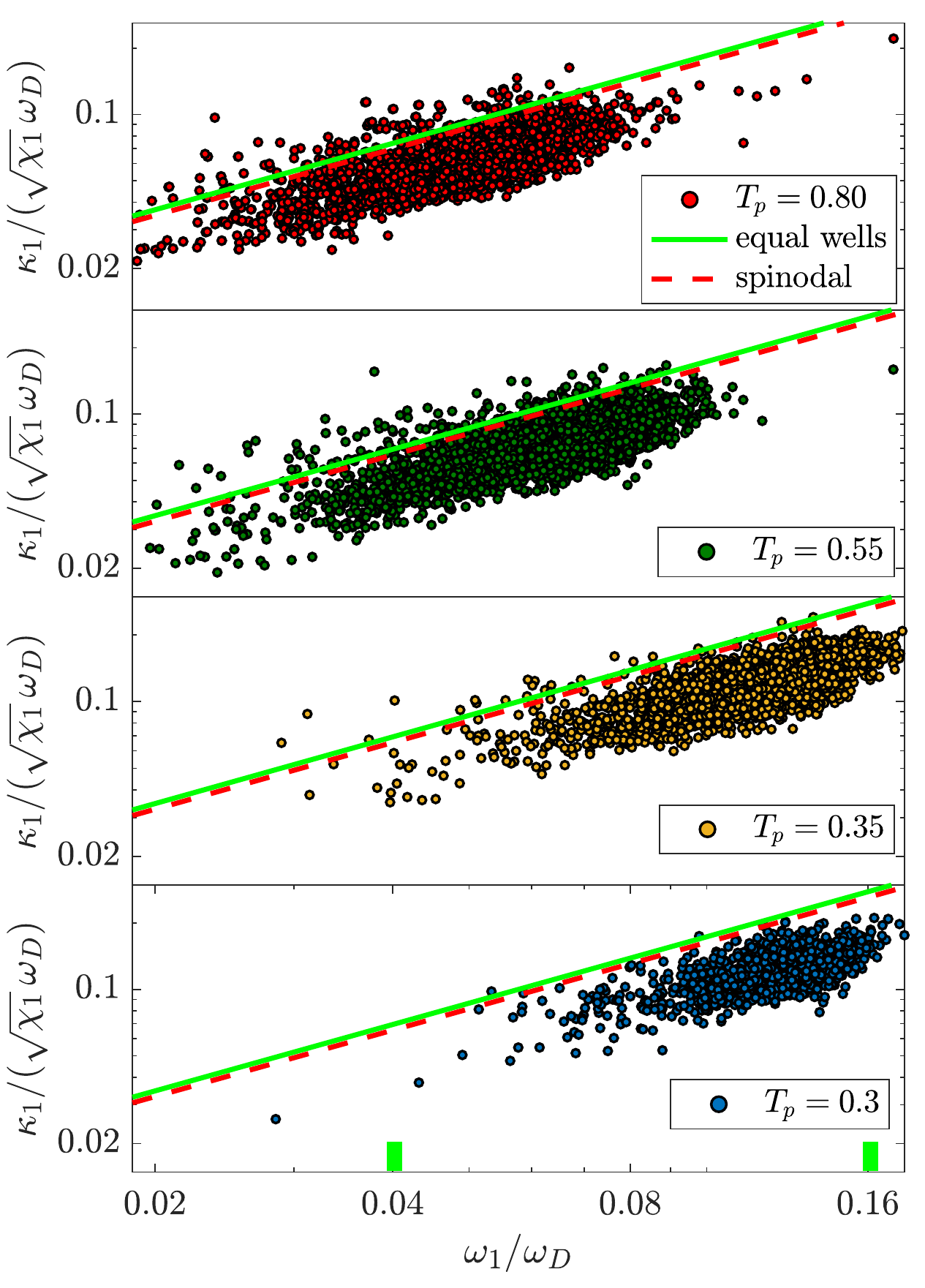}
	\caption{Scatter of $\kappa_1/(\sqrt{\chi}\omega_D)$ vs.~$\omega_1/\omega_D$ for non-linear modes at four $T_p$.  The green curve $\kappa_1/\sqrt{\chi}=\sqrt{3} \omega_1$ indicates the symmetric double wells and the red dashed curve $\kappa_1/\sqrt{\chi}=\sqrt{8/3}\omega_1$ indicates the spinodal case according to our model \cite{Wencheng19}. A higher proportion of dots below the red dashed line is found at low $T_p$.}
	\label{fig:phase_kappa}
\end{figure}

To estimate  $f(\omega_0)$, we Taylor expand local potentials along the non-linear modes (also called `cubic modes`)  \cite{Gartner16b,Kapteijns20} which capture the local energy landscapes better.  See SI for the detail of non-linear modes.   We thus get the coefficients $\omega_1$, $\kappa_1$, and $\chi_1$.  Note that $\chi_1$ (or $c_1$) is insensitive to the Taylor expansion along non-linear modes or normal modes (see SI).  Fig.~\ref{fig:phase_kappa} shows the scatter of $\kappa_1/(\sqrt{\chi_1}\omega_D) $ vs.~$\omega_1/\omega_D$ at four $T_p$ we have.  From Eq.~(\ref{f_w}), we first estimate the numerator $P(\omega_0,\kappa_0)$ by calculating the average density $P(\omega,\kappa)$ between the green line and the red dashed line in the range $\omega_1/\omega_D \in[0.04,0.16]$ (between two green bars) instead of at $\kappa_0$ and $\omega_0$. Then the denominator $\int_0^\infty P(\omega_0,\kappa)d\kappa$ in Eq.~(\ref{f_w}) is averaged in this $\omega_1$ range as well.  The green line $\kappa_1/\sqrt{\chi}=\sqrt{3} \omega_1$ corresponds to the two wells with equal depth and the red dashed line $\kappa_1/\sqrt{\chi}=\sqrt{8/3}\omega_1$ corresponds to the spinodal case according to our model \cite{Wencheng19}. See the estimated $f(\omega_0)$ in Table~\ref{tabn0}. We tested that the order of magnitude of  $f(\omega_0)$ will not change  if we choose a narrower range of $\omega_1$ at three higher $T_p$ .   At the lowest $T_p=0.3$,  we only find two points in the integral interval and $f(\omega_0)$ strongly depends on the range we choose.  However,  the estimations of $D_L(\omega_0)$ and $C_0$ are reliably and  they are smaller at lower $T_p$, and they ensure the decrease of $n_0$.  The dramatic decrease in $f(\omega_0)$ because of the higher proportion of single wells (dots below the spinodal line in Fig.~\ref{fig:phase_kappa}) is consistent with the recent findings that DWPs are rarer than QLMs in ultrastable glasses \cite{Kapteijns20,Khomenko21}.

\renewcommand\arraystretch{2}	
	\begin{table}[htb!]

		\begin{centering}
			\begin{tabular}{|c|c|c|c|c|c|c|c|}
				\hline 	
				$T_{p}$& $\omega_0/\omega_{D}$ & $C_{0}[(\rm{Jm{^3}s)^{-1}}]$ & $D_{L}(\omega_0)[\rm{s}]$ & $f(\omega_0)$& $n_{0}\rm[J^{-1}m^{-3}]$ \tabularnewline
				\hline 
				\hline 
				$0.80$ & $0.06$ &  $5.4\times10^{62}$ & $5.3\times10^{-15}$ & $0.36$& $1.0\times10^{48}$\tabularnewline
				\hline 
				$0.55$ & $0.07$ &  $4.8\times10^{62}$ & $3.8\times10^{-16}$ & $0.26$& $4.7\times10^{46}$\tabularnewline
				\hline 
				$0.35$ & $0.94$ &  $3.7\times10^{62}$ & $2.5\times10^{-16}$ & $0.05$& $4.2\times10^{45}$\tabularnewline
					\hline 
				$0.30$ & $0.10$ & $3.5\times10^{62}$ & $5.9\times10^{-17}$ & $0.03$& $6.1\times10^{44}$\tabularnewline
				\hline 
			\end{tabular}
			\par\end{centering}
		
		\caption{ The estimations of $\omega_0$, $C_0$, $D_L(\omega_0)$, and $f(\omega_0)$ at four $T_p$ in amorphous silicon. The density of TLS $n_0=C_0D_L(\omega_0)f(\omega_0)$  decreases dramatically with lowering $T_p$.}
			\label{tabn0}  
	\end{table}

From the above estimations, $C_0$ is insensitive to the system preparation. Both $D_L(\omega_0)$ and $f(\omega_0)$ decrease significantly with lowering $T_p$ which leads to the significant decrease in $n_0$ by  a factor of over $1000$.  The estimated $n_0$ agrees with the experimental measurement of $n_0$ in amorphous silicon which varies between $10^{45}-10^{48}$ $\rm{J^{-1}m^{-3}}$  \cite{Queen13}. 
\\

\paragraph{Conclusion}
We have built up the quantitative relationship between the density of TLS and the density of QLMs. We found that (i) TLS which contribute to the thermal transport, correspond to the  QLMs with frequency $\omega_0$ that is about $5\%$ to $10\%$ of $\omega_D$; 
(ii) The decrease in the density of TLS $n_0$ is not only influenced by the rarefaction of the density of QLMs  $D_L(\omega_0)$ but also influenced by the decrease of $f(\omega_0)$.  The latter reflects the distribution of local energy landscapes; (iii) The estimations of $n_0$ are consistent with that found in experiments in amorphous silicon.  Although we relied on several  assumptions and even some of them are inaccurate (see the discussion below), we think that for the order of magnitude of $n_0$, our method is effective since we grasped the properties of key parameters $\omega_0$, $D_L(\omega_0)$ and $f(\omega_0)$ in $n_0$.  

The previous work  \cite{Parshin07} argues that TLS are dominated by the lower bound $\Delta_{\min}$, which corresponds to the longest experimental time $\tau$, to explain some universality in glasses. However, we think that TLS are dominated  by the tunneling whose time scale is much smaller than $\tau$.  Because of the successful estimations on the dramatic decrease in $n_0$ and the right order of $\omega_0/\omega_D$ that is supported by the recent work \cite{Khomenko21} where a more careful calculation for the tunneling  is considered,  our picture appears to be closer to the prediction of TLS in experiments  \cite{Queen13}.

The estimated $n_0$ in our approach in most stable glasses is more than $100$ times smaller than that in \cite{Khomenko20}. 
Indeed, the quartic potentials (Eq.(\ref{U})) and nonlinear modes we adopted do not accurately describe the DWPs according to their numerical results, but we think it will not change the  order of magnitude of $n_0$.  One may think that they  overestimated $n_0$ by using the classical thermal activations to probe the DWPs since most of TLS they found corresponding to long tunneling times (lower bound of $\Delta_0$) will not be found in experiments. However, it is not the case since the order of magnitude of $n_0$ is insensitive to changing the lower bound of $\Delta_0$ according to our argument. The main source of discrepancy of $n_0$ might be that  the system they used is different (particle density, pair interaction) which leads to that $A_4$ \cite{Wang19} varies less significantly than the Ref.~\cite{Rainone20} that we used.

It is worth noting that, besides Ref.\cite{Khomenko20,Khomenko21}, some other works \cite{Demichelis99,Reinisch04,Reinisch05,Damart18,Mizuno20} also probe the DWPs directly, and some of them look at the properties of DWPs along the MEPs to estimate $n_0$ numerically. Clearly, we can get $n_0$ from Eq.(\ref{n0}) once we get $P(\omega_0,\kappa_0)$ through MEPs. Due to lack of simulation technology on MEPs, we cannot compare  $n_0$ in this way to our results. 

 Admittedly, in our adopted approach the estimation on $f(\omega_0)$ is not precise,  but we believe that  $f(\omega_0)$ decreasing with lowering $T_p$ should be correct, and that, at least at high $T_p$,  $f(\omega_0)$ we estimated is a good approximation.  $f(\omega_0)$ would be proportional to  the ratio of $C_{\rm TLS} /C_{sm}$, where $C_{\rm TLS}\propto n_0$ and $C_{sm}\propto D_L(\tilde{\omega})$ with typical $\tilde{\omega}$ are the prefactors of  specific heat contributions for TLS and local harmonic oscillators, respectively. This ratio is indeed found to be smaller in experiments in more stable amorphous material  $\rm B_2O_3$ \cite{Ramos04}.

In the latest work \cite{Wencheng21}, the geometry of  low-energy local excitations (local rearrangements) is systemically studied where  the local excitations become more localized in more stable glasses. It would be interesting to verify it on tunneling particles \cite{Khomenko20}. If that is true,  then the first-principles calculations, for example, Path Integral Molecular Dynamics method \cite{Marx96},  in real materials would be possible, which might reveal the entities of TLS in the end.

To explain the rarefaction of QLMs (dramatic decrease of $A_4$),  Ref. \cite{Wencheng20} proposes a thermal excitations picture that is controlled by a typical frequency, and Ref. \cite{Rainone21} uses an interacting harmonic oscillators model with some controlling parameters.  But a third model (or picture) using the parameters that can be measured directly in computer glasses to explain the behavior of $A_4$  quantitatively is needed.   Since $A_4$ reflects the level of the stability of glasses, the understanding of $A_4$ would shed light on developing a new method to  prepare   more stable computer glasses in the future.
\\

%----------------------------
\paragraph{Acknowledgmts.}

We thank E.~Bouchbinder, A.~Kumar, I.~Procaccia, T.W.J.~de Geus, F.~Zamponi, and Y.~Zheng for useful discussions, and M.~Wyart for useful discussions in the whole process. We thank F.~Zamponi for some specific suggestion.  We thank the authors
of \cite{Rainone20} for providing us with equilibrated swap configurations. We also thank the Simons Collaboration on Cracking the Glass Problem which provides a great platform to  exchange knowledge and form the initial idea of this work.
%%_____________________________________________________________

%\balance
\bibliographystyle{unsrtnat}
\bibliography{library_tls}
\bibliographystyle{apsrev4-1}
%_____________________________________________________________
\onecolumngrid

\begin{center}
	\rule{200pt}{0.5pt}
\end{center}

\appendix
\newpage

\clearpage
\begin{center}
	\medskip
	\textbf{Supporting information --} \textbf{ Toward understanding the depletion of two-level systems in ultrastable glasses}
\end{center}

\vspace{2em}
\section{ Calculation of the Jacobian $\left\Vert \frac{d\omega\wedge d\kappa}{dE\wedge d\Delta_0}\right\Vert $}

To calculate the Jacobian $\left\Vert \frac{d\omega\wedge d\kappa}{dE\wedge d\Delta_{0}}\right\Vert $, we first calculate the energy difference $\delta\varepsilon$. For
a double-well potential, $U(s)=\frac{1}{2}m\omega^{2}s^{2}+\frac{\kappa}{6}s^{3}+\frac{1}{24}\chi s^{4}$,
the positions of two minima are $s_{0}=0$ and $s_{0}=\frac{-3\kappa-3\sqrt{\kappa^{2}-8m\omega^{2}\chi/3}}{2\chi}$.
The energy difference of two minima is

\begin{align}
	\delta\varepsilon & =\frac{\left(3\kappa+\sqrt{9\kappa^{2}-24m\omega^{2}\chi}\right)^{2}\left(-3\kappa^{2}+12m\omega^{2}\chi-\kappa\sqrt{9\kappa^{2}-24m\omega^{2}\chi}\right)}{192\chi^{3}}.
\end{align}
Its derivative with respect to $\kappa$ at $\kappa_{c}$ is $\frac{d\delta\varepsilon}{d\kappa}|_{\kappa=\kappa_{c}}=\frac{4\kappa_{c}^{3}}{3\chi^{3}}$,
where $\kappa_{c}=\sqrt{3m\chi}\omega$ (the symmetric double well
condition). We also know $\delta\varepsilon=\sqrt{E^{2}-\Delta_{0}^{2}}$
and $\Delta_{0}=W\exp\left(-\left(\frac{\omega}{\bar{\omega}}\right)^{3}\right)$.

Then the Jacobian $\left\Vert \frac{d\omega\wedge d\kappa}{dE\wedge d\Delta_{0}}\right\Vert $
in the vicinity of symmetric wells is:

\begin{align}
	\left\Vert \frac{d\omega\wedge d\kappa}{dE\wedge d\Delta_{0}}\right\Vert  & =\left|\frac{dE}{d\kappa}\,\frac{d\Delta_{0}}{d\omega}\right|^{-1}\nonumber \\
	& =\left|\frac{dE}{d\delta\varepsilon}\frac{d\delta\varepsilon}{d\kappa}\,\frac{d\Delta_{0}}{d\omega}\right|^{-1}\nonumber \\
	& \simeq\left[\frac{\delta\varepsilon}{E}\frac{4\kappa_{c}^{3}}{3\chi^{3}}\,\frac{3W}{\omega}\left(\frac{\omega}{\bar{\omega}}\right)^{3}\exp\left(-\left(\frac{\omega}{\bar{\omega}}\right)^{3}\right)\right]^{-1}\nonumber \\
	& =\frac{\chi^{3}\omega}{4\kappa_{c}^{3}}\left(\frac{\bar{\omega}}{\omega}\right)^{3}\frac{E}{\Delta_{0}\sqrt{E^{2}-\Delta_{0}^{2}}}\nonumber \\
	& =\frac{\kappa_{c}}{18\hbar}\left(\frac{\bar{\omega}}{\omega}\right)^{6}\frac{E}{\Delta_{0}\sqrt{E^{2}-\Delta_{0}^{2}}}
\end{align}

\section{Estimation of the Debye frequency at different $T_p$}
Numerically, we can estimate the Debye frequency by the formula $\omega_D(T_p) = \left[18 \pi^2 \rho / \left( 2 c_t^{-3} + c_l^{-3} \right)\right]^{1/3}$. $\rho \equiv N / V$ is the particle number density which is a constant; 
$c_t = \sqrt{G / (m \rho)}$ and $c_l = \sqrt{(B + 4 G/3 )/ (m \rho)}$
are the transverse and longitudinal velocity,
related to the shear modulus $G$ and bulk modulus $B$;
$m$ is the particle mass (taken equal for all particles). Once we set to $\omega_D(T_p=0.55)=530k_{B}K/\hbar$, we can calculate $\omega_D$ by
\begin{align*}
	\omega_{D}(T_{p}) & =\frac{530k_{B}K}{\hbar}\left[\frac{2G^{-3/2}+\left(B+4G/3\right)^{-3/2}}{2G_{0}{}^{-3/2}+\left(B_{0}+4G_{0}/3\right)^{-3/2}}\right]^{-1/3}
\end{align*}
where $B_0$ and $G_0$ are the values at $T_p=0.55$. We do not need to worry the units of $B$ (or $G$) since they are canceled.
We get  $\omega_D(T_p=0.3)\approx610k_{B}K$, $\omega_D(T_p=0.35)\approx590k_{B}K$, and $\omega_D(T_p=0.8)\approx500k_{B}K$.

\section{Extract $A_4$ from $D_L(\omega)$ for small $\omega$}
\begin{figure} [hbt!]
	\centering
	\includegraphics[width=0.5\linewidth]{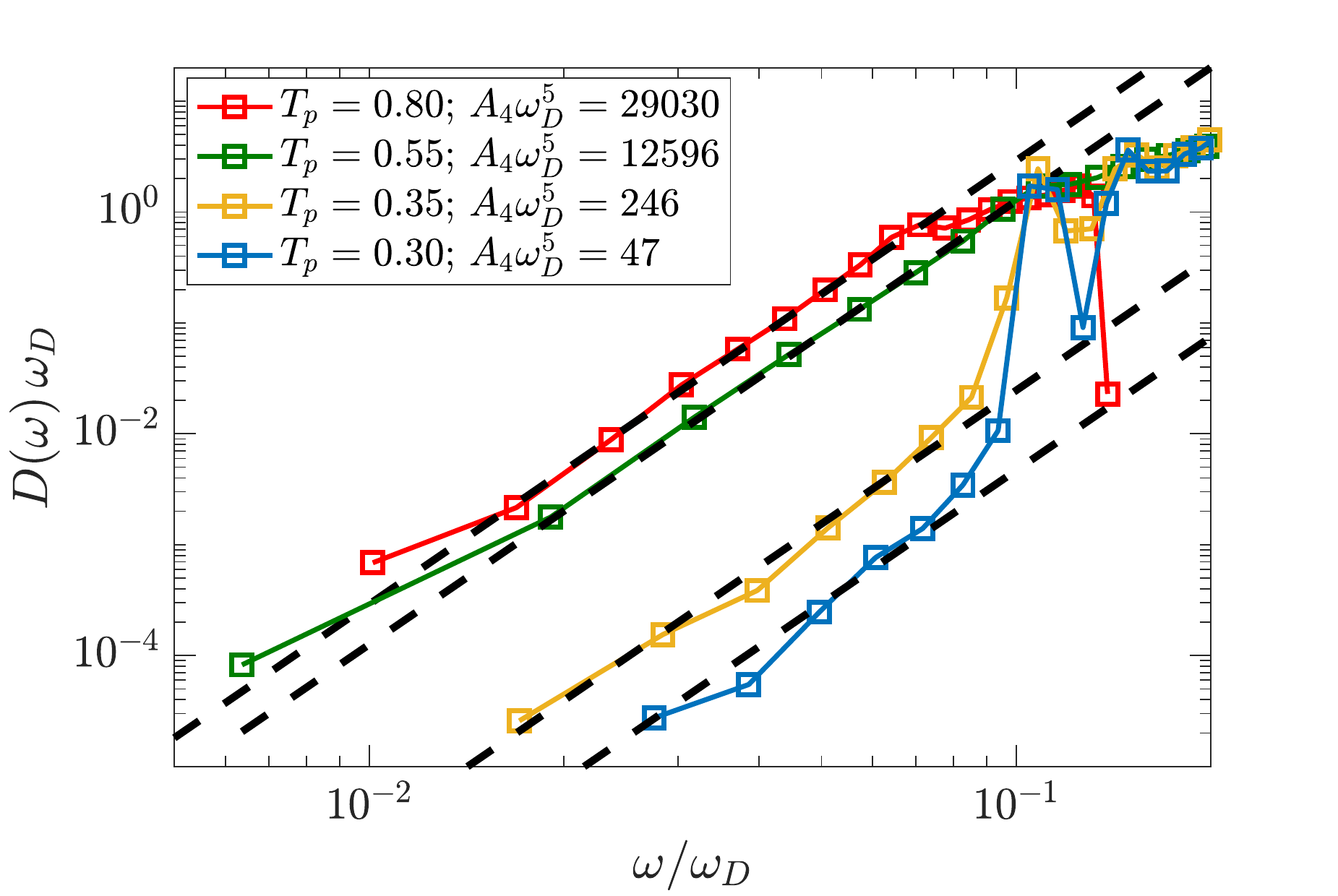}
	\caption{Density of normal modes $D(\omega)\omega_D$ vs $\omega/\omega_D$. $D(\omega)=D_L(\omega)=A_4\omega^4$ for small $\omega$. $D(\omega)$ is calculated by $D(\omega) \equiv\left\langle\frac{1}{3 N} \sum_{i} \delta\left(\omega-\omega_{i}\right)\right\rangle$, where $\langle\cdot\rangle$ stands for the ensemble average and $\omega_{i}$ is the discrete eigenfrequency of the Hessian matrix. 
	}
	\label{fig:Dw}
\end{figure}

We show the density of QLMs $D_L(\omega)\omega_D$ vs.~$\omega/\omega_D$ (both of which are dimensionless) for small $\omega$ in Fig.~\ref{fig:Dw}.  Since$\int D(\omega)\,d\omega=1$, we have $\int D(\omega)\,\omega_{D}\,d\left(\omega/\omega_{D}\right)=1$.
For small $\omega$, $D(\omega)\omega_{D}=A_{4}\omega_{D}^{5}\left(\omega/\omega_{D}\right)^4$
where the prefactor $A_{4}\omega_{D}^{5}$ is dimensionless as well.
The black dashed lines indicate the fitted  $D(\omega)\omega_{D}$ for small $\omega$.

\section{Non-linear modes}
\label{Ach4_3}
The non-linear modes $\vec{z}_f$ can be obtained by minimizing the cost function by applying the steepest descent method that starts from the QLMs directions. The cost function \cite{Gartner16b,Kapteijns20} is  
\begin{align*}
	F=\frac{\left|U^{(3)}\bullet\vec{z}\vec{z}\vec{z}\right|^{2}}{\left|H\bullet\vec{z}\vec{z}\right|^{3}}
\end{align*}
where $z$ is initially along the QLMs directions. Note that $F$ does not depend on the magnitude of $z$ and $F$ is proportional to the energy barrier when the system is close to the instability. The total interaction $U=\sum_{i}\sum_{j>i}\varphi$ where $\varphi$ is the pair interaction potential.   
$U^{(3)}=\frac{\partial^{3}U}{\partial\vec{r}\partial\vec{r}\partial\vec{r}}$ and $H=\frac{\partial^{2}U}{\partial\vec{r}\partial\vec{r}}$ where $H$ is the Hessian matrix. $\bullet$ means the contraction. Specifically

\begin{align}
	H\colon\vec{z}\vec{z}= \sum_{i<j}\left[\left(\frac{\varphi''}{\left|r_{ij}\right|^{2}}-\frac{\varphi'}{r_{ij}^{3}}\right)\left(\vec{r}_{ij}\cdot\vec{z}_{ij}\right)^{2}+\frac{\varphi'}{r_{ij}}\left(\vec{z}_{ij}\cdot\vec{z}_{ij}\right)\right]
\end{align}

\begin{align}
	U^{(3)}\bullet\vec{z}\vec{z}\vec{z}=  \sum_{i<j}\left\{ \left(\frac{\varphi^{'''}}{\left|r_{ij}\right|^{3}}-3\frac{\varphi^{''}}{\left|r_{ij}\right|^{4}}+3\frac{\varphi'}{\left|r_{ij}\right|^{5}}\right)\left(\vec{r}_{ij}\cdot\vec{z}_{ij}\right)^{3}+\left(\frac{\varphi^{''}}{\left|r_{ij}\right|^{2}}-\frac{\varphi'}{\left|r_{ij}\right|^{3}}\right)3\left(\vec{r}_{ij}\cdot\vec{z}_{ij}\right)\left(\vec{z}_{ij}\cdot\vec{z}_{ij}\right)\right\} 
\end{align}
$\varphi^{'''}$ ($\varphi^{''}$, $\varphi^{'}$)  is third (second, first) derivative of the pair interaction with respect to $r_{ij}$. The coefficients $\lambda_1,\kappa_1,\chi_1$ thus are obtained by the Taylor expansion of the potential $U$ along $z_f$.  Nonlinear modes in Fig.~2 in main text are obtained from QLMs whose frequencies are below the frequency of the first plane waves.

\section {Estimation of $c_1$}
\label{Ach4_4}
We show the scatter plot of ${\chi_1} a^2/m\omega_D^2$ ($\chi a^2/m\omega_D^2$) v.s.~$\omega_1^2$ ($\omega^2$) for non-linear modes (QLMs) at four $T_p$. $\chi_1$ is similar to $\chi$ for small $\omega$ at each $T_p$. We estimate $c_1$ from  non-linear modes by taking the median value of $\chi_1 a^2/m\omega_D^2$. Specifically, $c_1\approx 0.5,\,0.8,\,2.1,\,2.5$ from high to low $T_p$. The median of $\chi a^2/m\omega_D^2$ obtained from QLMs for the lowest $50$ softest QLMs (left side of the vertical lines)  is approximately equal to $0.3,\,0.6,\,2.4,\,2.3$.

\begin{figure}[hbt!]
	\centering
	\includegraphics[width=0.8\linewidth]{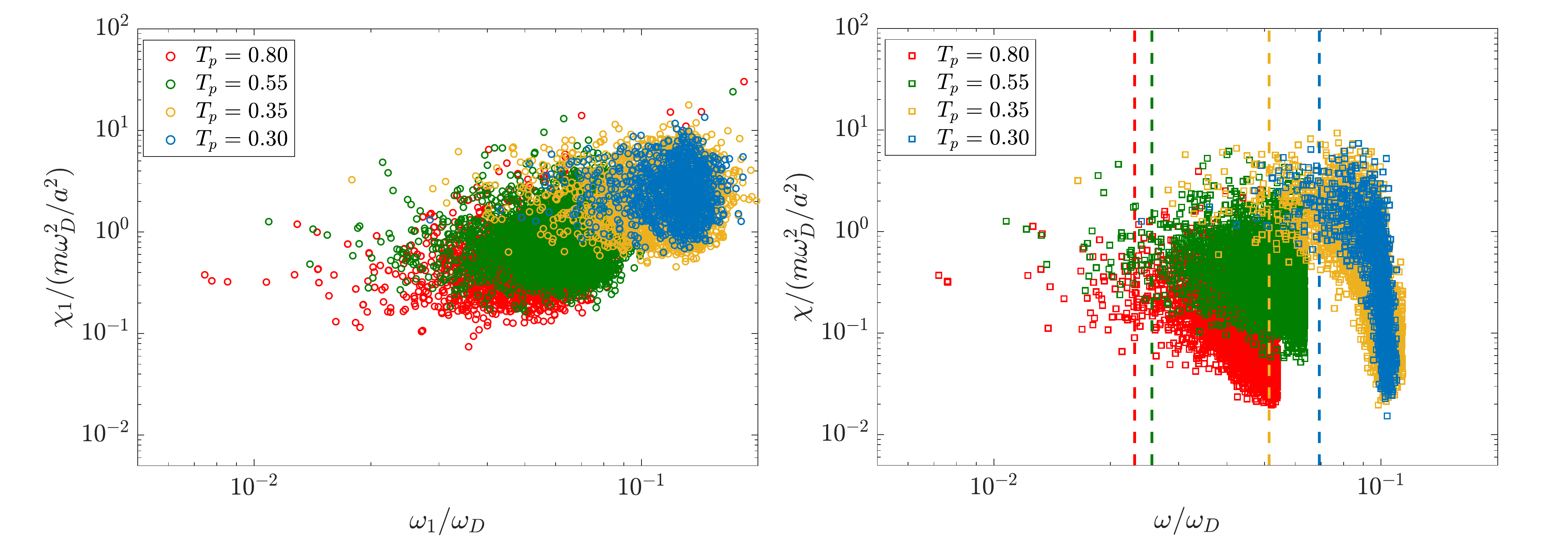}
	
	\caption{Scatter plots of $\chi_1 a^2/m\omega_D^2$ vs.~$\omega_1/\omega_D$ (left) and $\chi a^2/m\omega_D^2$ v.s. $\omega/\omega_D$ (right). For big $\omega$, $\chi$ becomes smaller because QLMs are hybridized with plane waves. The vertical lines are the thresholds to estimate the median of $\chi a^2/m\omega_D^2$. }
	\label{fig:chi2}
\end{figure}

\end{document}